\documentclass[aps,prl,twocolumn,superscriptaddress,showpacs]{revtex4-1}
\usepackage{graphicx}
\usepackage{dcolumn}
\usepackage{bm}
\usepackage{subfigure}
\usepackage{amsmath}
\usepackage{amssymb}
\usepackage{hyperref} 
\usepackage{xcolor}

\newcommand{\ua}{\uparrow} 
\newcommand{\da}{\downarrow}
\newcommand{\ra}{\rightarrow}

\newcommand{\SRO}{Sr$_2$RuO$_4$} 
 
\usepackage{times} 

\begin{document}
\title{Hidden anomalous Hall effect in \SRO~with chiral superconductivity dominated by the Ru $d_{xy}$ orbital}
\author{Jia-Long Zhang}
\address{Kavli Institute for Theoretical Sciences, University of Chinese Academy of Sciences, Beijing 100190, China}
\address{Shenzhen Institute for Quantum Science and Engineering \& Guangdong Provincial Key Laboratory of Quantum Science and Engineering, Southern University of Science and Technology, Shenzhen 518055, Guangdong, China}
\author{Yu Li}
\address{Kavli Institute for Theoretical Sciences, University of Chinese Academy of Sciences, Beijing 100190, China}
\author{Wen Huang}
\email{huangw3@sustech.edu.cn}
\address{Shenzhen Institute for Quantum Science and Engineering \& Guangdong Provincial Key Laboratory of Quantum Science and Engineering, Southern University of Science and Technology, Shenzhen 518055, Guangdong, China}
\author{Fu-Chun Zhang}
\email{fuchun@ucas.ac.cn}
\address{Kavli Institute for Theoretical Sciences, University of Chinese Academy of Sciences, Beijing 100190, China}
\address{Center for Excellence for Topological Quantum Computation, Chinese Academy of Sciences, Beijing 100190, China}

\date{\today}

\begin{abstract}
The polar Kerr effect in superconducting \SRO~implies finite ac anomalous Hall conductivity. Since intrinsic anomalous Hall effect (AHE) is not expected for a chiral superconducting pairing developed on the single Ru $d_{xy}$ orbital, multiorbital chiral pairing actively involving the Ru $d_{xz}$ and $d_{yz}$ orbitals has been proposed as a potential mechanism. Here we propose that AHE could still arise even if the chiral superconductivity is predominantly driven by the $d_{xy}$ orbital. This is demonstrated through two separate models which take into account subdominant orbitals in the Cooper pairing, one involving the oxygen $p_x$ and $p_y$ orbitals in the RuO$_2$ plane, and another the $d_{xz}$ and $d_{yz}$ orbitals. In both models, finite orbital mixing between the dominant $d_{xy}$ and the other orbitals may induce inter-orbital pairing between them, and the resultant states support intrinsic AHE, with Kerr rotation angles that could potentially reconcile with the experimental observation. Our proposal therefore sheds new light on the microscopic pairing in \SRO. We also show that intrinsic Hall effect is generally absent for non-chiral states such as $\mathcal{S}+i\mathcal{D}$, $\mathcal{D}+i\mathcal{P}$ and $\mathcal{D}+i\mathcal{G}$, which provides a clear constraint on the symmetry of the superconducting order in this material. 
\end{abstract}

\maketitle
{\bf Introduction.} --The nature of the unconventional superconducting pairing in \SRO~is an outstanding open question in condensed matter physics. Despite tremendous efforts on various fronts, it remains difficult and controversial to interpret all of the key experimental observations in a consistent theory~\cite{Maeno:94,Maeno:01,Mackenzie:03,Kallin:09,Kallin:12,Maeno:12,Liu:15,Kallin:16,Mackenzie:17}. A number of measurements point to time-reversal symmetry breaking (TRSB) pairing -- indicative of the condensation of multiple superconducting order parameters~\cite{Luke:98,Xia:06,Grinenko:20}, most likely in the two-dimensional irreducible representations (irrep) of the underlying crystalline $D_{4h}$ group~\cite{Mackenzie:03}. This makes chiral pairings, including chiral p-wave ($\mathcal{P}_x+i\mathcal{P}_y$) and chiral d-wave ($\mathcal{D}_{xz}+i\mathcal{D}_{yz}$), promising candidates, although mixed-representation states, such as the chiral d-wave with $\mathcal{D}_{x^2-y^2}+i\mathcal{D}_{xy}$, and the non-chiral states the likes of $\mathcal{S}+i\mathcal{D}$ and $\mathcal{D}+i\mathcal{P}$, etc, cannot be definitively ruled out. The chiral pairings support chiral edge modes, and the p-wave state may further support Majorana zero modes that could be utilized for topological quantum computing~\cite{Kitaev:03,Nayak:08}.

One important evidence for TRSB pairing in this material is the Kerr rotation, i.e. a circularly polarized light normally incident on a superconducting sample is reflected with a rotated polarization~\cite{Xia:06}. To date, the origin of the Kerr effect, or that of the closely related ac anomalous Hall effect (AHE) in superconducting \SRO, remains controversial. It has often been inquired alongside the question about the primary superconducting orbital(s) in this material. To begin with, a single-orbital chiral pairing, as would be the case if superconductivity is solely associated with the quasi-two-dimensional (2D) Ru $d_{xy}$ orbital, is not expected to generate anomalous Hall response as a consequence of Galilean invariance~\cite{Read:00,Yip:92,Roy:08,Lutchyn:08}, except in the presence of impurities~\cite{Goryo:08,Lutchyn:09,Konig:17,LiYu:19}. However, such an extrinsic mechanism may not be sufficient to explain the Kerr rotation in measurements on high-quality crystals~\cite{Kallin:09,Goryo:08,Lutchyn:09}. Recent discussions about possible active pairing on the two quasi-one-dimensional (1D) Ru $d_{xz}$ and $d_{yz}$ orbitals have stimulated an alternative interpretation, that intrinsic AHE is not forbidden in such a multiorbital chiral superconductor~\cite{Taylor:12,Wysokinski:12,Komendova:17}. 

The Kerr effect bears special significance for understanding the microscopic Cooper pairing in \SRO, including its driving superconducting orbital(s) and the symmetry of its superconducting order parameter. Our paper contributes new insights into both of these two highly contentious issues.

On the one hand, we disclose a `hidden' AHE in pristine \SRO~even when its chiral pairing is dominated by the single Ru $d_{xy}$ orbital, i.e. the quasi-2D $\gamma$-band. We are motivated by the observation of substantial mixing between the $d_{xy}$ orbital and the oxygen $p_x$ and $p_y$ orbitals in the RuO$_2$ plane. Although they locate relatively far from the Fermi energy in the atomic limit and are thus customarily ignored in most theoretical studies, the oxygen orbitals in fact contribute significantly to the $\gamma$-band density of states~\cite{Vaugier:12,Oguchi:95}. We will show that, despite having only one band crossing the Fermi energy, such system with a hidden multiorbital character exhibits intrinsic Hall response and will hence generate Kerr rotation. The effect is found to rely crucially on the induced inter-orbital pairing between the Ru-$d$ and O-$p$ orbitals. The conclusion applies to all chiral superconducting states, including chiral p-wave, as well as chiral d-wave with $\mathcal{D}_{xz}+i\mathcal{D}_{yz}$ and $\mathcal{D}_{x^2-y^2}+i\mathcal{D}_{xy}$ pairings. In a simple generalization, qualitatively similar Hall response is obtained in another model containing the three $t_{2g}$ orbitals, where finite orbital mixing between the dominant $d_{xy}$ and the subdominant $d_{xz}/d_{yz}$ orbitals may similarly induce inter-orbital pairings. 

On the other hand, our study also places strong constraints on the possible superconducting pairing symmetry in this material. In particular, we show that non-chiral TRSB states such as $\mathcal{S}+i\mathcal{D}$, $\mathcal{D}+i\mathcal{P}$ and $\mathcal{D}+i\mathcal{G}$ generally do not support intrinsic Hall response, irrespective of the microscopic model details.  

\begin{figure}
\includegraphics[width=8cm]{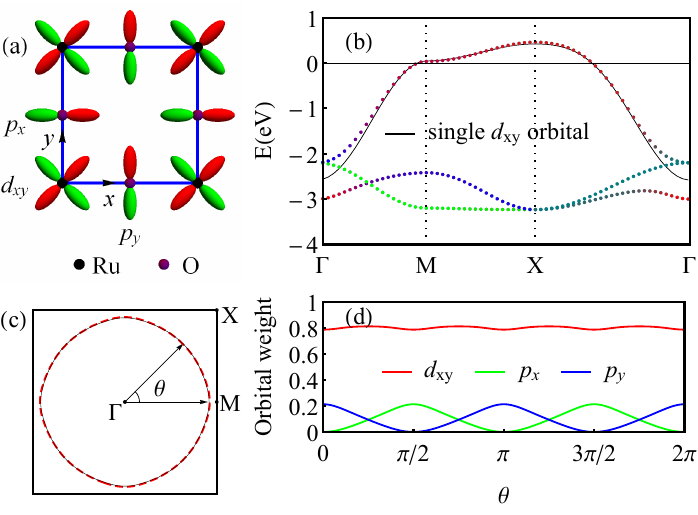}
\caption{Lattice and electronic structure of the $dpp$-model. (a) Sketch of the $dpp$-model with Ru-$d_{xy}$ and O-$p_x$ and $p_y$ orbitals on the square lattice of the RuO$_2$ plane. (b) Band structure of the $dpp$-model (dotted lines) and the single-orbital model with $d_{xy}$ orbital (solid line). We choose a set of tight-binding parameters to match with the first-principle calculations~\cite{Vaugier:12}. The color gradient in dotted lines encrypts the variation of the weights of the three orbitals in the electronic states. The color codes are shown in (d). (c) Fermi surfaces of the $dpp$ (red dashed) and the single-$d_{xy}$-orbital-model (black solid). (d) Angular dependence of the individual orbital weights across the Fermi surface.}
\label{fig:fig1}
\end{figure}

{\bf Chiral superconductivity in $dpp$-model.}-- We start by constructing a tight-binding model consisting the Ru $d_{xy}$ and O $p_{x}$ and $p_y$ orbitals in the RuO$_2$ plane (Fig.~\ref{fig:fig1}), which we name the $dpp$-model. The normal state Hamiltonian is given by $H_0 =\sum_{\vec{k}\sigma} \psi^\dagger_{\vec{k}\sigma}\hat{H}_{0\vec{k}}\psi_{\vec{k}\sigma}$, where $\sigma=\ua,\da$ is the spin index, $\psi_{\vec{k}\sigma}=(c_{d\vec{k}\sigma},c_{p_x\vec{k}\sigma},c_{p_y\vec{k}\sigma})^{\intercal}$ represents the fermionic spinor, and
\begin{equation}
\hat{H}_{0\vec{k}}=\begin{pmatrix}
\epsilon_{d\vec{k}} & i t_{dp} \sin \frac{k_y}{2} &  it_{dp} \sin \frac{k_x}{2} \\
-i t_{dp} \sin \frac{k_y}{2} & \epsilon_{p_x\vec{k}} & t_{pp} \sin \frac{k_x}{2} \sin \frac{k_y}{2} \\
-i t_{dp} \sin \frac{k_x}{2} &  t_{pp} \sin \frac{k_x}{2} \sin \frac{k_y}{2} & \epsilon_{p_y\vec{k}}
\end{pmatrix} \,,
\label{eq:H0}
\end{equation}
Here $\epsilon_{d\vec{k}}=-2t_d(\cos k_x+\cos k_y)-4t^\prime_d \cos k_x \cos k_y -\mu_d$, $\epsilon_{p_{x(y)}\vec{k}}=-2t_{p\parallel(\perp)}\cos k_x -2t_{p\perp(\parallel)} \cos k_y -\mu_p$. Here $t_d$ and $t_d^\prime$ stand respectively for the first and second neighbor hoppings between $d_{xy}$ orbitals, and $t_{\parallel(\perp)}$ denotes the first neighbor hopping of the $p$ orbitals parallel (perpendicular) to the orbital's lobe direction. It is worth noting that, due to spatial proximity, the $d$-$p$ mixing $t_{dp}$ is among the largest hopping integrals in the model.

Throughout this section, we employ the following set of parameters $(t_d, t^\prime_d, t_{p\parallel}, t_{p\perp}, t_{dp}, t_{pp}, \mu_d, \mu_p)=(0.35, 0.14, -0.25, 0.074, 1, 0.33, 1.04, 2.55)$ in unit of eV, which leads to the band structure and Fermi surface as shown in Fig.~\ref{fig:fig1} (b) and (c). This band structure, with a band inversion between the $d$ and $p$ orbitals at the $\Gamma$-point, shows good agreement with the one obtained from first-principle calculations~\cite{Vaugier:12}. Due to the very band inversion, the resultant $\gamma$-band defies an effective single-Wannier-orbital construction. Crucial to our argument, the $p$ orbitals are found to feature prominently at the Fermi energy [Fig.~\ref{fig:fig1} (d)], in total representing roughly twenty-percent of the electronic density of states. An early band structure calculation also found similar-size contribution from the $p$ orbitals~\cite{Oguchi:95}. This observation would otherwise raise an interesting question about the role played by the oxygen orbitals in the microscopic theories of the superconductivity in \SRO.

In the following, we illustrate the construction of the gap functions of the $dpp$-model with the example of the spin-triplet chiral p-wave pairing. Similar construction for several other pairing states is presented in the Supplementary~\cite{Supplementary}. The chiral p-wave order parameter belongs to the $E_u$ irrep of the $D_{4h}$ point group, and as per our assumption, the pairing is dominated by the $d_{xy}$ orbital, with $\Delta_{d\vec{k}}=\Delta_1(\sin k_x +i\sin k_y)$. Although the $p$ orbitals are distant from the Fermi energy in the unhybridized limit and thus likely do not exhibit intrinsic Cooper instability~\cite{footnote1}, the strong $d$-$p$ mixing may still induce some inter-orbital pairing under appropriate circumstances. Note that, for simplicity, throughout the paper the gap functions are presumed to have the forms of the simplest lattice harmonics. No qualitative feature is lost due to this simplification. Furthermore, since the model in Eq.~(\ref{eq:H0}) ignores spin-orbit coupling (SOC), components of the $E_u$ irrep with $\sin k_z$-like pairings~\cite{Huang:18} are ignored.

Before turning to the forms of the inter-orbital pairing, an important remark about the gap classification is in order. While the usual intra-orbital pairing is fully classified according to the spin exchange statistics and the spatial parity of the Cooper pair wavefunction, inter-orbital pairings are also characterized by the parity number under orbital exchange (orbital-singlet vs orbital-triplet). Furthermore, in analysing the symmetry of the pairing, the spatial parity of the individual electron orbitals constituting the Cooper pair also matters. This is because symmetry operations act on both the Cooper pair wavefunction and the orbital wavefunctions of the two constituent electrons~\cite{Huang:19b,Ramires:19,Kaba:19}. This additional degree of freedom adds a layer of complexity when pairing takes place between electron orbitals of opposite parities, such as the $d$ and $p_{x(y)}$ orbitals in the present study. In short, the superconducting gap functions could acquire forms that differ considerably from the lattice generalizations of $k_x+ik_y$. The same holds true for all other irreps~\cite{Huang:19b,Ramires:19,Kaba:19}. As a consequence, each irrep may permit multiple coexisting gap functions coupled by orbital mixing~\cite{Huang:19b}.

Since the orbitals $d_{xy}$ and $(p_x,p_y)$ belong respectively to the $B_{2g}$ and $E_u$ irreps, the pair creation (or annihilation) operators $c^\dagger_{d}c^\dagger_{p_y}$ and $c^\dagger_{d}c^\dagger_{p_x}$ jointly transform according to the $E_u$ irrep. Denoting the inter-orbital pairing function between the $d_{xy}$ and $p_{x(y)}$ orbitals $\Delta_{dp_{x(y)}\vec{k}}$, their lowest order basis functions are then $1+a k_x^2 + b k_y^2$ and $1+b k_x^2 + a k_y^2$, where $a$ and $b$ are real constants. Accounting for the lattice structure in Fig.~\ref{fig:fig1} (a), one may take $\Delta_{dp_{x(y)}\vec{k}} =\Delta_{2} \cos \frac{k_{y(x)}}{2}$. In contrast to the intra-orbital (spatial) odd-parity gap function $\Delta_{d\vec{k}}=\Delta_1(\sin k_x + i\sin k_y)$, the inter-orbital odd-parity pairing features even-parity gap functions, while its oddness is encoded instead in the electron orbital manifold, i.e. the product of $d$ and $p$ orbital wavefunctions being odd under inversion.

The full pairing term of our model then follows as, $\sum_{\vec{k},\sigma\neq\bar{\sigma}}\psi^\dagger_{\vec{k}\sigma}\hat{\Delta}_{\vec{k}}(\psi^\dagger_{-\vec{k},\bar{\sigma}})^{\intercal} + h.c.$, where,
\begin{equation}
\hat{\Delta}_{\vec{k}}=\begin{pmatrix}
\Delta_{d\vec{k}} & e^{i\alpha}\Delta_{dp_x\vec{k}} & e^{i\beta}\Delta_{dp_y\vec{k}} \\
-e^{i\alpha}\Delta_{dp_x\vec{k}} & 0 & 0 \\
-e^{i\beta} \Delta_{dp_y\vec{k}} & 0 & 0
\end{pmatrix} \,,
\label{eq:gapfunction1}
\end{equation}
Here the gap amplitudes $\Delta_1$ and $\Delta_2$ are taken to be real and positive, and the phases $\alpha$ and $\beta$ remain to be determined. Notice that the inter-orbital pairings are orbital-singlet. To understand how inter-orbital mixing couples $\Delta_1$ and $\Delta_2$ and fixes $\alpha$ and $\beta$, we evaluate the second order term in the standard free energy expansion, $f^\text{2nd}= T\sum_l \sum_{\vec{k},w_m}\text{Tr}[\hat{g}(iw_m,\vec{k})\hat{\Delta}_{\vec{k}}\hat{\bar{g}}(iw_m,\vec{k})\hat{\Delta}^\dagger_{\vec{k}}]^{2l}/(2l)$, where $\hat{g}(iw_m,\vec{k})=(iw_m - \hat{H}_{0\vec{k}})^{-1}$ and $\hat{\bar{g}}(iw_m,\vec{k})=(iw_m + \hat{H}_{0,-\vec{k}}^\ast)^{-1}$ are the respective electron and hole components of the Gorkov Green's function, $T$ is the temperature, and $\omega_m=(2m+1)\pi T$ is the fermionic Matsubara frequency. This returns the following coupling term in the Ginzburg-Landau free energy,
\begin{equation}
f^\text{2nd}_\text{coupling} \approx \rho t_{dp} (\cos\alpha-\sin\beta)\Delta_1\Delta_2 \,,
\label{eq:D1}
\end{equation}
where $\rho$ is a real constant. The effect of the inter-orbital mixing becomes obvious: the energetically favorable choice would be $(\alpha,\beta)=(0,-\pi/2)$ if $\rho t_{dp}<0$, and $(\alpha,\beta)=(\pi,\pi/2)$ if $\rho t_{dp}>0$.

{\bf Hall conductivity and Kerr angle.}-- We now proceed to compute the Hall conductance of our model. Within linear response theory, it is given by the antisymmetric part of the current-current correlation function $\pi_{xy}(\vec{q},\omega)$,
\begin{equation}
\sigma_{H}(\omega)=\frac{i}{2\omega}\lim_{\vec{q}\rightarrow 0}\left[\pi_{xy}(\vec{q},\omega) -\pi_{yx}(\vec{q},\omega)\right],
\label{eq:sigma}
\end{equation}
where, at the one-loop approximation,
\begin{equation}
\begin{aligned}
\pi_{xy}(\vec{q}=0,i \omega_n)=&T\sum_{\vec{k},\omega_m}\text{Tr}\left[\hat{v}_{x\vec{k}} \hat{G}(\vec{k},i\omega_m)\right.\\
& \left. \times\hat{v}_{y\vec{k}} \hat{G}(\vec{k},i\omega_m+i\omega_n) \right],
\end{aligned}
\label{eq:oneloop}
\end{equation}
where $\omega_n=2n\pi T$ represents the bosonic Matsubara frequency, $\hat{v}_{x(y)\vec{k}}$ stands for the $x(y)$ component of the velocity operator and $\hat{G}(\vec{k},i\omega_m)=(i\omega_m-\hat{H}^{\text{BdG}}_{\vec{k}})^{-1}$ the full Green's function of the corresponding Bogoliubov de-Gennes Hamiltonian associated with Eqns. (\ref{eq:H0}) and (\ref{eq:D1}). In actual calculations we transform the Green's function into its spectral representation, and obtain an alternative form of Eq. (\ref{eq:oneloop}) that has also been employed in Ref.~\onlinecite{Gradhand:13}. The Hall conductivity is evaluated by using an analytical continuation to real frequencies $i\omega_n \ra  \omega+i\delta$, where $\delta$ is taken to be $10^{-5}$ throughout this study.

\begin{figure}
\includegraphics[width=8cm]{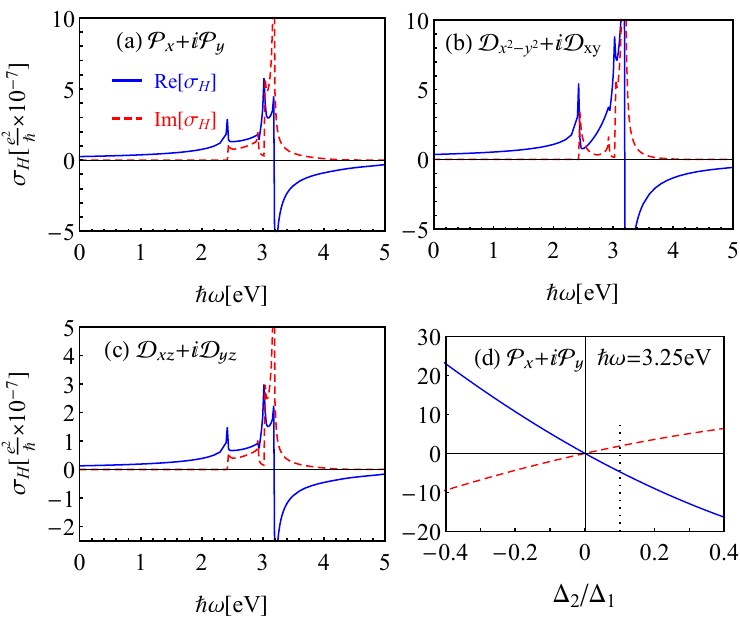}
\caption{The real (blue solid line) and imaginary (red dash line) part of the Hall conductivity for various chiral superconducting states at $T=0$: (a) $\mathcal{P}_x +i\mathcal{P}_y$, (b) $\mathcal{D}_{x^2-y^2}+i\mathcal{D}_{xy}$, (c)$\mathcal{D}_{xz}+i\mathcal{D}_{yz}$. The inter-orbital pairing between the $d_{xy}$- and the $p$-orbitals is set at one tenth of the intra-orbital pairing on $d_{xy}$. The gap functions of the latter two states are presented in the Supplementary~\cite{Supplementary}. (d) The $\Delta_2$-dependence of the Hall conductance at fixed $\Delta_1=0.35$ meV and $\hbar\omega=3.25$eV.}
\label{fig:fig2}
\end{figure}

Figure~\ref{fig:fig2} presents the representative numerical results for three different chiral states. In accord with a recent experimental estimate~\cite{Sharma:20}, we took $\Delta_1=0.35$meV. Without loss of generality, in Fig.~\ref{fig:fig2} (a), (b) and (c) we have taken inter-orbital pairings that are one tenth of the intra-orbital pairing on the $d_{xy}$ orbital. Given the substantial $d$-$p$ mixing, such a modest assumption may not be entirely unreasonable, as we substantiate in the Supplementary~\cite{Supplementary}. Overall, the Hall response of the three chiral states shows no qualitative difference. Quantitatively, it is interesting to note that the Hall conductivity in $\mathcal{P}_x+i\mathcal{P}_y$ is approximately twice as large as that in $\mathcal{D}_{xz}+i\mathcal{D}_{yz}$. The factor of $1/2$ arises from a $k_z$-integration involving the square of $\sin k_z$ associated with the latter's gap function. In contrast, these two states feature the same thermal Hall conductance~\cite{Yoshioka:18} and similar spontaneous surface current~\cite{Nie:19}.

As in previous multiorbital models~\cite{Taylor:12,Wysokinski:12,Gradhand:13,Wang:18,Brydon:19}, the intensity of $\text{Im}[\sigma_{H}]$ originates from the transitions between pairs of the states belonging to different branches of the Bogoliubov bands, one with positive energy $E_1$ and the other with negative $-E_2$ ($E_1 \neq E_2$ and $E_1,E_2>0$). The lower cutoff frequency $\omega_c$ at which $\text{Im}[\sigma_{H}]$ becomes none-zero is determined by the details of the model. For the specific set of parameters we use, $\hbar\omega_c \approx 2.5$eV and the corresponding onset intensity is associated with transitions near the M-point of the BZ.

It is worth stressing that a finite inter-orbital pairing is essential for the emergence of AHE for the pairing model given in Eq. (\ref{eq:gapfunction1}). Exemplified by the chiral p-wave state as shown in Fig.~\ref{fig:fig2} (d), both real and imaginary parts of $\sigma_{H}$ drop to zero linearly as the inter-orbital pairing dereases. To gain a better understanding, we derive the Hall conductivity of the reduced, and more analytically tractable models containing only $d_{xy}$ and $p_x$ (or $d_{xy}$ and $p_y$) orbitals. As shown in the Supplementary~\cite{Supplementary}, the reduced models have approximately $\sigma_{H} \propto \Delta_2$ for the chiral p-wave state in the limit of small $\Delta_2/\Delta_1$, consistent with the above numerical results.

Finally, to connect with the optical Kerr measurement, we evaluate the Kerr rotation angle given by
\begin{equation}
\theta_K = \frac{4\pi}{\omega l}\text{Im}\left[\frac{\sigma_{H}(\omega)}{n(n^2-1)} \right]\,,
\label{eq:KerrAngle}
\end{equation}
where $l$ stands for the interlayer spacing between the RuO$_2$ planes and $n$ the frequency-dependent refractive index. Following the estimate in Ref.~\onlinecite{Taylor:12}, at the experimental photon energy $\hbar \omega=0.8$eV we find $\theta_K \approx 27,40$ and $14$ nrads for the three respective chiral pairings in Fig.~\ref{fig:fig2}. They are not far off from the measured value at low temperatures~\cite{Xia:06}. Nonetheless, caution is needed when using these estimates at face value, as we lack an accurate prediction for the inter-orbital pairing strength $\Delta_{2}$.

{\bf $d^3$-model.}-- In the presence of finite SOC and interlayer coupling, the $d_{xy}$ orbital also mixes with the $d_{xz}$ and $d_{yz}$ orbitals. These couplings are much weaker than $t_{dp}$~\cite{Bergemann:03,Haverkort:08,Veenstra:14}. However, the two quasi-1D $d$ orbitals in fact lie closer in energy to $d_{xy}$ than the $p$ orbitals do in the $dpp$-model. Moreover, according to a recent theoretical calculation~\cite{Gingras:19}, sizable inter-orbital pairing between the $d$ orbitals is not entirely impossible. It is therefore sensible to consider a model containing all three $t_{2g}$ orbitals and with induced inter-orbital pairing between the dominant $d_{xy}$ and the other two orbitals (which we refer to as $d^3$-model). Note that our motivation differs from a previous study which had assumed comparable pairing instabilities on all three $t_{2g}$ orbitals~\cite{Gradhand:13}.

Figure \ref{fig:fig3} (b) presents the Hall conductance of a chiral p-wave state (see Supplementary~\cite{Supplementary} for details). In distinction to the $dpp$-model, the characteristic peaks of $\sigma_{H}(\omega)$ emerge at rather different frequencies due to a very different low energy quasiparticle spectrum. Notably, at $\hbar \omega=0.8$eV, we obtain $\theta_K \approx 63.6$ nrads under the modest assumption of $\Delta_2=\Delta_1/10$. This is again close to the experimental observation, and we expect similar qualitative behavior for other chiral states.

\begin{figure}[]
\centering
\includegraphics[width=8cm]{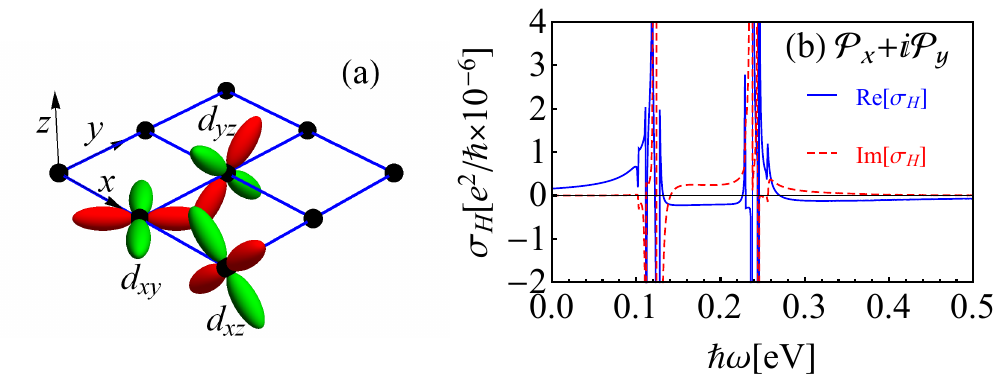}
\caption{(a) Lattice structure of the $d^3$-model. Each site of the square lattice hosts all three Ru $t_{2g}$ orbitals. (b) Real (blue solid line) and imaginary (red dash line) part of the Hall conductivity for a chiral p-wave state in the $d^3$-model. Details of the model are provided in the Supplementary~\cite{Supplementary}.}
\label{fig:fig3}
\end{figure}

{\bf Non-chiral states.}-- There have been frequent discussions of non-chiral TRSB orders in \SRO~\cite{Pustogow:19,Roising:19,Romer:19,Romer:20,Kivelson:20,Chronister:20,Scaffidi:20}. These states condense multiple superconducting order parameters belonging to distinct 1D irreps. However, we find that they most likely do not exhibit intrinsic Hall effect. Some of these states preserve certain vertical mirror symmetries. For this class, $\sigma_{H}$ as given by Eqs. (\ref{eq:sigma}) and (\ref{eq:oneloop}) exactly vanishes due to mutually cancelled contributions at any pair of $\vec{k}$'s related by the corresponding mirror reflections. This follows naturally from the symmetry property of the velocity operators under mirror reflections. Some examples are pairings of the forms $A_{1g}+iB_{1g}$ and $B_{1g}+iA_{2g}$ -- typically referred to as $\mathcal{S}+i\mathcal{D}_{x^2-y^2}$ and $\mathcal{D}_{x^2-y^2} + i \mathcal{G}_{xy(x^2-y^2)}$, respectively. Some mixed-parity states, for instance $A_{1g} +i A_{1u}$ (a mixture of s-wave and helical p-wave pairings, or simply $\mathcal{S}+i\mathcal{P}$), break all underlying vertical mirror symmetries. However, as we verify numerically, $\hat{v}_{x\vec{k}} \hat{G}(\vec{k})\hat{v}_{y\vec{k}}\hat{G}(\vec{k})-(x \leftrightarrow y)$ is zero at every $\vec{k}$, irrespective of the details of the underlying microscopic model conceivable for this material. Hence these states shall also exhibit vanishing Hall conductance. In short, the presence of a chiral superconducting order parameter appears to be critical for a Hall response to arise in pristine \SRO. This is consistent with the intuitive expectation by an analogy with quantum Hall insulators.

{\bf Concluding remarks.}-- While polar Kerr effect in ultra-clean \SRO~may rule against non-chiral states, it cannot reliably discriminate the various chiral states, as our study suggests. A final identification must then be made in conjunction with other key observations. For example, except in rare fine-tuned cases, the $\mathcal{P}_x+i\mathcal{P}_y$ and $\mathcal{D}_{xz}+i\mathcal{D}_{yz}$ states, or more precisely the chiral states in the $E_u$ and $E_g$ irreps, generically support finite spontaneous edge current~\cite{Huang:15,Nie:19}-- which however has eluded experimental detection~\cite{Kirtley:07,Hicks:10,Curran:14}. The $\mathcal{D}_{x^2-y^2}+i\mathcal{D}_{xy}$ state, on the other hand, is understood to produce vanishingly small surface current~\cite{Huang:14,Tada:15}. However, this state, formally classified as $B_{1g}+iB_{2g}$, shall typically exhibit two separate superconducting transitions, whereas experiments have only identified one~\cite{Yonezawa:14,Li:19}; symmetry analysis would also rule out discontinuities in all shear elastic moduli at the lower superconducting transition, yet a discontinuity was reported in the modulus $c_{66}$~\cite{Ghosh:20}. A final conclusion therefore still seems somewhat distant, and many experimental progresses are being made lately~\cite{Pustogow:19,Ishida:19,Sharma:20,Petsch:20,Ghosh:20,Benhabib:20,Chronister:20}.

In summary, although multiple electron orbitals and a pairing with chirality must both be involved for superconducting \SRO~to produce intrinsic AHE and Kerr rotation, the Ru quasi-1D $d_{xz}$ and $d_{yz}$ orbitals need not proactively participate in the Cooper pairing. Even when the pairing is driven solely by the Ru $d_{xy}$ orbital, inter-orbital pairing between this and other orbitals may emerge due to orbital mixing, which then leads to an intrinsic Hall effect. We have demonstrated this for two separate models, one taking into account the O $p_x$ and $p_y$ orbitals in the RuO$_2$-plane, and another the Ru $d_{xz}$ and $d_{yz}$ orbitals. We evaluated the corresponding Kerr rotation angle and made connection with the experimental measurement. In this light, it seems worthwhile to reassess the microscopic theories of the Cooper pairing and the question about the driving superconducting orbital(s) in this material~\cite{Agterberg:97,Mazin:99,Nomura:00,Nomura:02,Zhitomirsky:01,Ng:00,Eremin:02,RPA1,RPA2,Annett:06,RPA3,Raghu:10,Puetter:12,Huo:13,Wang:13,Kivelson:13,Yanase:14,Scaffidi:14,Tsuchiizu:15,Huang:16,Ramires:16,ZhangLD:18,WangWS:19,Romer:19,Roising:19,Gingras:19,WangZQ:20,Romer:20,Suh:19,Lindquist:19,ChenW:20,Scaffidi:20}. Finally, our proposal may be readily generalized to other TRSB superconductors where Kerr rotation has also been reported, including UPt$_3$, URu$_2$Si$_2$, PrOs$_4$Sb$_{12}$ and UTe$_2$~\cite{Schemm:14,Schemm:15,Levenson:18,Hayes:20}.

{\it Acknowledgments} -- We are grateful to Catherine Kallin, Aline Ramires and Zhiqiang Wang for their critical reading and comments on an earlier version of the manuscript. We also acknowledge fruitful discussions with Thomas Scaffidi, Manfred Sigrist, Qiang-Hua Wang and Fan Yang. This work is supported by NSFC under grant No.~11674278 (FCZ) and No.~11904155 (WH and JLZ), the strategic priority research program of CAS grant No. XDB28000000 (FCZ), and Beijing Municipal Science and Technology Commission Project No. Z181100004218001 (FCZ), the Guangdong Provincial Key Laboratory under Grant No.~2019B121203002 (WH), and the China Postdoctoral Science Foundation under Grant No. 2020M670422 (YL). Computing resources are provided by the Center for Computational Science and Engineering at Southern University of Science and Technology.

\setcounter{equation}{0}
\setcounter{figure}{0}
\renewcommand {\theequation} {S\arabic{equation}}
\renewcommand {\thefigure} {S\arabic{figure}}
\begin{widetext}

\section{Supplemental materials}
This supplementary contains four sections and is organized as follows: In Section A we present pairing functions for other chiral states as well as a non-chiral TRSB state in the $dpp$-model. Concerning the importance of inter-orbital pairing in our study, we then demonstrate in Section B how inter-orbital mixing may induce inter-orbital $d$-$p$ pairing in the presence of a dominant intra-orbital pairing on the $d_{xy}$ orbital. In Section C we discuss a reduced model containing the $d_{xy}$ and only one of the $p$ orbitals. In this case analytical result is obtainable and through which the role of inter-orbital pairing in bringing about the Hall effect in $dpp$-model can be explicitly seen. Finally in Section D we construct a model containing three $t_{2g}$ orbitals, i.e., $d^3$-model.

\subsection{A. Superconducting gap functions in the $dpp$-model}
In the maintext we have constructed the chiral p-wave pairing in the $dpp$-model. Here, we shall present the gap functions of the other chiral states as well as a non-chiral TRSB state. In the main text, we have kept the constructed tight-binding model in two spatial dimensions. However, for the superconducting pairings, we will also consider three-dimensional ones. Note that, since the system preserves inversion symmetry, pairings of opposite parities cannot mutually induce one another. In like manner, the absence of SOC suggests that spin-singlet pairing cannot induce spin-triplet pairing, or vice versa. Nevertheless, such pairings could still coexist in the presence of appropriate electron interactions in the respective pairing channels.

(i). $\mathcal{D}_{x^2-y^2}+i\mathcal{D}_{xy}$: This is an even-parity pairing with a linear mixture of order parameters in the $B_{1g}$ and $B_{2g}$ irreps of the $D_{4h}$ point group. The intra-orbital pairing on the $d_{xy}$ orbital is spin-singlet in nature, and we assume its gap function takes the form $\Delta_1 (\cos k_x - \cos k_y) + i\Delta_1^\prime\sin k_x\sin k_y$. Here, $\Delta_1$ and $\Delta_1^\prime$ represent the respective amplitudes of the $B_{1g}$ and $B_{2g}$ components, and they are in general different in magnitude. According to the symmetry analysis in the maintext, the inter-orbital pairing between the $d$ and $p$ orbitals acquire the following forms in the small-$k$ representation,
\begin{eqnarray}
B_{1g}:~~~~\Delta_2 \left( k_y c^\dagger_{d\vec{k}\ua}c^\dagger_{p_x,-\vec{k}\da}- k_x c^\dagger_{d\vec{k}\ua}c^\dagger_{p_y,-\vec{k}\da} \right) \nonumber  \,, \\
B_{2g}:~~~~\Delta^\prime_2 \left( k_x c^\dagger_{d\vec{k}\ua}c^\dagger_{p_x,-\vec{k}\da}+ k_y c^\dagger_{d\vec{k}\ua}c^\dagger_{p_y,-\vec{k}\da} \right) \,,
\label{eq:chiralD}
\end{eqnarray}
where again the two gap amplitudes $\Delta_2$ and $\Delta_2^\prime$ are in general different. These are spin-singlet and orbital-singlet pairings, meaning that exchanging the spin indices or the orbital labels changes the sign of the gap functions. Upon spatial inversion, $k_i \ra -k_i$ and $c^\dagger_{p_{x(y)}} \ra -c^\dagger_{p_{x(y)}}$ whereas $c^\dagger_{d} \ra c^\dagger_{d}$, hence these pairings are also even-parity. The simplest lattice generalization appropriate for our model is then given by replacing $k_i$ with $\sin\frac{k_i}{2}$.

\begin{equation}
\hat{\Delta}_{\vec{k}} =\begin{bmatrix}
\Delta_{1}(\cos k_x -\cos k_y)+i\Delta_{1}^\prime\sin k_x \sin k_y & e^{i\phi}(\Delta_{2}\sin\frac{k_y}{2} +i\Delta_2^\prime\sin\frac{k_x}{2}) & e^{i\phi}(-\Delta_{2}\sin\frac{k_x}{2} +i\Delta_2^\prime\sin\frac{k_y}{2}) \\
-e^{i\phi}(\Delta_{2}\sin\frac{k_y}{2} +i\Delta_2^\prime\sin\frac{k_x}{2}) & 0 & 0 \\
-e^{i\phi}(-\Delta_{2}\sin\frac{k_x}{2} +i\Delta_2^\prime\sin\frac{k_y}{2}) & 0 & 0
\end{bmatrix} \,.
\end{equation}
Following the Ginzburg-Landau analysis in the maintext, we arrive at the 2nd order coupling between the order parameters,
\begin{equation}
f^\text{2nd}_\text{coupling} \sim \rho t_{dp} \sin\phi \Delta_1\Delta_2 + \rho^\prime t_{dp} \sin\phi \Delta^\prime_1\Delta^\prime_2 \,.
\end{equation}
We thus see that, depending on the sign of $\rho$ and $\rho^\prime$ and on the relative magnitude of the products $\Delta_1\Delta_2$ and $\Delta^\prime_1\Delta^\prime_2$, $\phi$ must be chosen between $\pi/2$ and $-\pi/2$ to minimize the free energy. For our purpose, it does not lose generality to take $\phi=\pi/2$. Further, besides taking $\Delta_2 = \Delta_1/10$ and $\Delta_2^\prime = \Delta_1^\prime/10$, we also assume $\Delta_1^\prime=\Delta_1/2$ in the numerical calculations.

(ii). $\mathcal{D}_{xz}+i\mathcal{D}_{yz}$: This is a spin-singlet even-parity pairing in the $E_g$-irrep. The intra-orbital pairing on the $d_{xy}$-orbital takes the form $\Delta_1(\sin k_x + i\sin k_y)\sin k_z$. The corresponding basis of the inter-orbital pairings between the $d$- and $p$-orbitals are given by,
\begin{equation}
E_u: ~~~~(k_z c^\dagger_{d\vec{k}\ua}c^\dagger_{p_x\vec{k}\da}, k_z c^\dagger_{d\vec{k}\ua}c^\dagger_{p_y\vec{k}\da}) \,.
\end{equation}
As in the previous case, these are also even-parity, spin-singlet and orbital-singlet pairings. A general gap function then follows as,
\begin{equation}
\hat{\Delta}_{\vec{k}}=\begin{bmatrix}
\Delta_1(\sin k_x+i\sin k_y)\sin k_z & e^{i\alpha}\Delta_2 \cos\frac{k_y}{2}\sin k_z & e^{i\beta}\Delta_2 \cos\frac{k_x}{2}\sin k_z \\
-e^{i\alpha}\Delta_2 \cos\frac{k_y}{2}\sin k_z & 0 & 0 \\
-e^{i\beta}\Delta_2 \cos\frac{k_x}{2}\sin k_z & 0 & 0
\end{bmatrix} \,.
\end{equation}
Following the free-energy analysis as in the maintext, $(\alpha,\beta)$ must be either $(0,\pi/2)$ or $(\pi,-\pi/2)$ to minimize the free energy.

(iii). $\mathcal{D}_{x^2-y^2}+i\mathcal{S}$: In the following we present the gap function of a representative non-chiral state, $\mathcal{D}_{x^2-y^2}+i\mathcal{S}$, which is a mixture of $A_{1g}$ and $B_{1g}$ components. The intra-orbital pairing on the $d_{xy}$-orbital is spin-singlet and can be expressed in the simple form $\Delta_1 (\cos k_x - \cos k_y) + i\Delta_1^\prime$, while the inter-orbital pairings acquire the following forms,
\begin{eqnarray}
B_{1g}:~~~~\Delta_2 \left( k_y c^\dagger_{d\vec{k}\ua}c^\dagger_{p_x,-\vec{k}\da}- k_x c^\dagger_{d\vec{k}\ua}c^\dagger_{p_y,-\vec{k}\da} \right) \,, \nonumber \\
A_{1g}:~~~~\Delta^\prime_2 \left( k_y c^\dagger_{d\vec{k}\ua}c^\dagger_{p_x,-\vec{k}\da}+ k_x c^\dagger_{d\vec{k}\ua}c^\dagger_{p_y,-\vec{k}\da} \right) \,.
\label{eq:DpIS}
\end{eqnarray}
Following the free-energy analysis and without loss of generality, one may write,
\begin{equation}
\hat{\Delta}_{\vec{k}} =\begin{bmatrix}
\Delta_{1}(\cos k_x -\cos k_y)+i\Delta_{1}^\prime & i(\Delta_{2}\sin\frac{k_y}{2} +i\Delta_2^\prime\sin\frac{k_y}{2}) & i(-\Delta_{2}\sin\frac{k_x}{2} +i\Delta_2^\prime\sin\frac{k_x}{2}) \\
-i(\Delta_{2}\sin\frac{k_y}{2} +i\Delta_2^\prime\sin\frac{k_y}{2}) & 0 & 0 \\
-i(-\Delta_{2}\sin\frac{k_x}{2} +i\Delta_2^\prime\sin\frac{k_x}{2}) & 0 & 0
\end{bmatrix} \,.
\end{equation}
Other non-chiral states may be constructed in a similar manner, which we will not elaborate here.

\subsection{B. Induced inter-orbital pairing}
We demonstrate in this section how inter-orbital mixing may induce inter-orbital $d$-$p$ pairing in the presence of a dominant intra-orbital pairing on the $d_{xy}$ orbital by using the self-consistent mean-field BdG study of the $dpp$-model. This is best illustrated by the evolution of the $d$-$p$ pair correlation in the presence of a varying inter-orbital hybridization.

We consider the effective interaction on the $d$-orbitals in a particular Cooper channel, e.g.
\begin{equation}
H^d_\text{int} = -U_d\sum_{\vec{k},\vec{k}^\prime} \zeta_{\vec{k}}  \zeta_{\vec{k}^\prime}^\ast c^\dagger_{d,\vec{k}\ua}c^\dagger_{d,-\vec{k}\da} c_{d,-\vec{k}^\prime\da}c_{d,\vec{k}^\prime\ua} \,,
\end{equation}
where $\zeta_{\vec{k}}$ represents the form factor of the pairing gap function under consideration. A mean-field decoupling returns $ \sum_{\vec{k}}\left( \Delta_1\zeta_{\vec{k}}c^\dagger_{d,\vec{k}\ua}c^\dagger_{d,-\vec{k}\da} + h.c.  \right) $, where $\Delta_1 = -U_d/V  \sum_{\vec{k}^\prime}\langle \zeta_{\vec{k}^\prime}^\ast c_{d,-\vec{k}^\prime\da} c_{d,\vec{k}^\prime\ua} \rangle$ with $\langle . \rangle$ denoting the expectation value. The corresponding intra-orbital pair correlation is given by $\Delta_1/U_d$. Without assuming any interaction between the $d$ and $p$ orbitals, we self-consistently determine the $\Delta_1$ at a fixed $U_d$ and evaluate the pair correlation between the $d$ and $p$ orbitals. This correlation is given by
\begin{equation}
C_{dp} = \frac{1}{2V}\sum_{\vec{k}}\left( \langle \chi^x_{\vec{k}} c_{d,-\vec{k}\da}c_{p_x,\vec{k}\ua} \rangle + \langle \chi^y_{\vec{k}} c_{d,-\vec{k}\da}c_{p_y,\vec{k}\ua} \rangle \right) \,,
\end{equation}
where $\chi^{x(y)}_{\vec{k}}$ denote the form factors of the pairing between $d$ and $p_{x(y)}$ orbitals in the same symmetry channel, with appropriate phase factors encapsulated. For example, in the chiral p-wave model we have $\zeta_{\vec{k}} = \sin k_x + i\sin k_y$, $\chi_{\vec{k}}^{x} = e^{i\alpha}\cos \frac{k_{y}}{2}$ and $\chi_{\vec{k}}^{y} = e^{i\beta}\cos \frac{k_{x}}{2}$ according to Eq.~(\ref{eq:gapfunction1}). The following discussions will be based on our calculations on the chiral p-wave state, but the conclusion applies to all superconducting symmetries.

Figure \ref{fig:fig4} shows the relative strength between the inter-orbital and intra-orbital pair correlations as a function of the average $p$-orbital weight on the Fermi level. There is an apparent positive correlation between $C_{dp}$ and the $p$-orbital weight on the Fermi level. Crucially, this correlation converges in the weak-coupling limit. Therefore an inter-orbital $d$-$p$ pairing order parameter ($\Delta_2$ in the maintext) develops, provided there exists an effective attractive interaction $U_{dp}$ between the $d$ and $p$ orbitals in the corresponding superconducting channel. With appropriate relative phases, the intra-orbital and inter-orbital superconducting order parameters mutually enhance each other, as suggested in Eq.~(\ref{eq:D1}). At present, there is no microscopic prediction for the above effective interactions. Nonetheless, we take note of the tendency to develop a sizable inter-orbital pairing order parameter in the scenario of realistic $d$-$p$ mixing, i.e. around twenty percent $p$-orbital weight at the Fermi level. We have further checked that, even if the inter-orbital interaction $U_{dp}$ by itself cannot drive a stable superconducting pairing, a finite $\Delta_2$ readily develops once the intra-orbital interaction $U_d$ is turned on. In our numerical evaluation of the Hall conductivity in the maintext, we have taken a modest inter-orbital pairing $\Delta_2$ at one tenth of $\Delta_1$.

\begin{figure}
\includegraphics[width=8cm]{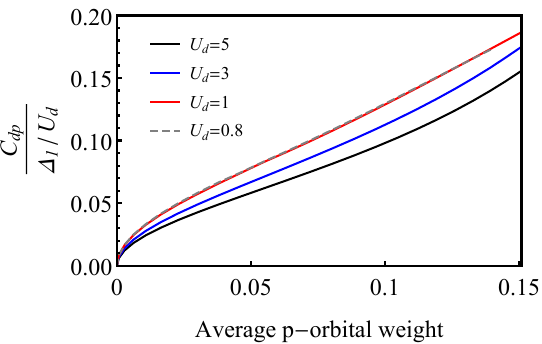}
\caption{Chiral $p$-wave state: the ratio between the inter-orbital $d$-$p$ and intra-orbital $d$-$d$ pair correlations, as a function of the average $p$-orbital weight on the Fermi surface. The $p$-orbital weight is tuned by varying the inter-orbital hybridization $t_{dp}$. To avoid drastic changes in the Fermi surface as $t_{dp}$ varies, we have chosen $\mu_d=-1$ and $\mu_p=6$, different than the ones given in the maintext. Other parameters are the same as given in the maintext. Note we have taken inter-orbital interaction $U_{dp}=0$ in these calculations.}
\label{fig:fig4}
\end{figure}

\subsection{C. Leading contributions to $\sigma_{H}$ in $dpp$-model}
To illustrate how the inter-orbital pairing is crucial in bringing about the Hall effect in our $dpp$-model, we analytically derive $\sigma_{H}$ for a reduced version of the model containing the $d_{xy}$ and only one of the $p$ orbitals. Such a two-orbital model is obtained simply by discarding terms in the Hamiltonian that are associated with the excluded $p$ orbital. This simplification is possible under our assumption of negligible intra-orbital pairing on the $p$ orbitals, as well as much weaker inter-orbital $d$-$p$ pairings compared to the intra-orbital pairing on the $d_{xy}$ orbital.

The Bogoliubov-de Gennes (BdG) Hamiltonian for the two-orbital $dp$
model in question can be written as $H^{\text{BdG}}=\sum_{\vec{k},\sigma}%
\psi_{\vec{k}\sigma}^{\dagger}\hat{H}_{\vec{k}}^{\text{BdG}}%
\psi_{\vec{k}\sigma}$, in which $\sigma=\uparrow$, $\downarrow$ is the spin
index, and $\psi_{\vec{k}\sigma}=( c_{d,\vec{k}\sigma
},c_{d,-\vec{k}\bar{\sigma}}^{\dagger},c_{p,\vec{k}\sigma}%
,c_{p,-\vec{k}\bar{\sigma}}^{\dagger} )^{\intercal}$ represents the
two-orbital Numbu spinor (for convenience, we have rearranged the spinor basis with respect to that in the maintext), and $\hat{H}_{\vec{k}}^{\text{BdG}}$ can expanded
in terms of Pauli matrices $\left\{  \hat{\tau}_{i}\right\}  \left(
i=1,2,3\right)  $ in particle-hole space as
\begin{equation}
\hat{H}_{\vec{k}}^{\text{BdG}}=\left(
\begin{array}
[c]{cccc}
\epsilon_{d\vec{k}} & \Delta_{d,\vec{k}} & t_{dp,\vec{k}} & \Delta
_{dp,\vec{k}}\\
\Delta_{d,\vec{k}}^{\ast} & -\epsilon_{d,-\vec{k}} & \Delta_{pd,\vec{k}}^{\ast} & -t_{dp,-\vec{k}}^{\ast}\\
t_{dp,\vec{k}}^{\ast} & \Delta_{pd,\vec{k}} & \epsilon_{p\vec{k}} &
0\\
\Delta_{dp,\vec{k}}^{\ast} & -t_{dp,-\vec{k}} & 0 & -\epsilon_{p,-\vec{k}}
\end{array}
\right) \,,
\end{equation}
where $\epsilon_{d\vec{k}}=-2t_{d}\left(  \cos k_{x}+\cos k_{y}\right)
-4t_{d}^{\prime}\cos k_{x}\cos k_{y}-\mu_{d}$, $\epsilon_{p_{x\left(  y\right)
}\vec{k}}=-2t_{p||\left(  \perp\right)  }\cos k_{x}-2t_{p\perp\left(
||\right)  }\cos k_{y}-\mu_{p}$ are the dispersions of the $d_{xy}$ and
$p_{x\left(  y\right)  }$ orbitals respectively, and $t_{dp_{x(y)},\vec{k}}=it_{dp}\sin\frac{k_{y\left(  x\right)  }}{2}$ is the
inter-orbital hybridization. This expression distinguishes between $\Delta_{pd,\vec{k}}$ and $\Delta_{dp,\vec{k}}$ which shall acquire the same form, except that $\Delta_{pd,\vec{k}} = -\Delta_{dp,\vec{k}}$ if the pairing is an orbital-singlet.

The velocity or current operators are obtained through a standard minimal coupling $-i\vec{ \nabla} \ra -i\vec{ \nabla}+\vec{ A}$ (we have set the charge $e=1$ for brevity) and then by taking the functional derivative with respect to the vector potential $\vec{ A}$. Note that since the momentum $\vec{ k}$ in the pairing function does not amount to center-of-mass motion of the Cooper pairs, $\Delta_{\vec{ k}}$ is not involved in the minimal coupling. In full momentum space, we arrive at,
\begin{align}
\hat{\upsilon}_{\vec{k}}^{i}  &  \equiv\left.  \frac{\delta\hat
{H}_{\vec{k}}^{\text{BdG}}\left[  \vec{A}\right]  }{\delta A_{i}%
}\right\vert _{\vec{A}=0}=\frac{\delta}{\delta A_{i}}\left(
\begin{array}
[c]{cccc}%
\epsilon_{d,\vec{k}+\vec{A}} & \Delta_{d,\vec{k}} & t_{dp,\vec{k}+\vec{A}} &
\Delta_{dp,\vec{k}}\\
\Delta_{d,\vec{k}}^{\ast} & -\epsilon_{d,-\vec{k}+\vec{A}} & \Delta_{pd,\vec{k}%
}^{\ast} & -t_{dp,-\vec{k}+\vec{A}}^{\ast}\\
t_{dp,\vec{k}+\vec{A}}^{\ast} & \Delta_{pd,\vec{k}} & \epsilon_{p,\vec{k}+\vec{A}} &
0\\
\Delta_{dp,\vec{k}}^{\ast} & -t_{dp,-\vec{k}+\vec{A}} & 0 & -\epsilon_{p,-\vec{k}+\vec{A}}%
\end{array}
\right)  _{\vec{A}=0}=\left(
\begin{array}
[c]{cc}%
\frac{\partial\epsilon_{d\vec{k}}}{\partial k_{i}} & \frac{\partial
t_{dp,\vec{k}}}{\partial k_{i}}\\
\frac{\partial t_{dp,\vec{k}}^{\ast}}{\partial k_{i}} & \frac{\partial
\epsilon_{p\vec{k}}}{\partial k_{i}}%
\end{array}
\right) \otimes \hat{\tau}_{0} \nonumber\\
&  \equiv \left(
\begin{array}
[c]{cc}%
\hat{\upsilon}_{d,\vec{k}}^{i} & \hat{\upsilon}_{dp,\vec{k}}^{i}\\
\hat{\upsilon}_{dp,\vec{k}}^{i\ast} & \hat{\upsilon}_{p,\vec{k}}^{i}%
\end{array}
\right) \otimes \hat{\tau}_{0},
\end{align}
where $i=\allowbreak x$, $y$ and $\hat{\tau}_0$ is a rank-2 identity matrix.

For the case orbital-singlet inter-orbital pairing $\Delta_{dp,\vec{k}}=-\Delta_{pd,\vec{k}}$ (which is the case for the multiple states constructed in the preceding section), one obtains the Hall conductivity following Eqns. (\ref{eq:sigma}) and (\ref{eq:oneloop}) and after a lengthy calculation,
\begin{equation}
\sigma_{H}^{dp}\left(  \omega\right)  =-2\sum_{\vec{k}}%
\frac{i\left[  \left(  \boldsymbol{\upsilon}_{d,\vec{k}}%
-\boldsymbol{\upsilon}_{p,\vec{k}}\right)  \times\boldsymbol{\upsilon
}_{dp,\vec{k}}\right]  _{z}\epsilon_{p,\vec{k}}\operatorname{Re}(
\Delta_{d,\vec{k}}^{\ast}\Delta_{dp,\vec{k}})  }{E_{+,\vec{k}%
}E_{-,\vec{k}}\left(  E_{+,\vec{k}}+E_{-,\vec{k}}\right)  \left[
\left(  E_{+,\vec{k}}+E_{-,\vec{k}}\right)  ^{2}-\left(  \omega
+i\delta\right)  ^{2}\right]  },
\label{eq:Sigmadp}
\end{equation}
where $E_{\pm,\vec{k}}>0$ denote the quasiparticle spectra. They satisfy the following relation,
\begin{align}
E_{\pm,\vec{k}}^{2}  &  =\frac{1}{2}\left(  \epsilon_{d\vec{k}}^{2}%
+\epsilon_{p\vec{k}}^{2}+2\left\vert t_{dp,\vec{k}}\right\vert ^{2}%
+\left\vert \Delta_{d,\vec{k}}\right\vert ^{2}+2\left\vert \Delta
_{dp,\vec{k}}\right\vert ^{2}\right)  \pm\frac{1}{2}\left\{  \left(
\epsilon_{d\vec{k}}^{2}+\epsilon_{p\vec{k}}^{2}+2\left\vert t_{dp,\vec{k}%
}\right\vert ^{2}+\left\vert \Delta_{d,\vec{k}}\right\vert ^{2}%
+2\left\vert \Delta_{dp,\vec{k}}\right\vert ^{2}\right)  ^{2}\right.
\nonumber\\
&  \left.  -4\left[  \left(  \epsilon_{d\vec{k}}\epsilon_{p\vec{k}}-\left\vert
t_{dp,\vec{k}}\right\vert ^{2}\right)  ^{2}+2\left(  \epsilon_{d\vec{k}}%
\epsilon_{p\vec{k}}+\left\vert t_{dp,\vec{k}}\right\vert ^{2}\right)
\left\vert \Delta_{dp,\vec{k}}\right\vert ^{4}+\epsilon_{p\vec{k}}%
^{2}\left\vert \Delta_{d,\vec{k}}\right\vert ^{2}+\left\vert
\Delta_{dp,\vec{k}}\right\vert ^{4}+4\epsilon_{p\vec{k}}\operatorname{Re}%
\text{(}\Delta_{d,\vec{k}}^{\ast}\Delta_{dp,\vec{k}}t_{dp,\vec{k}%
}\text{)}\right]  \right\}  ^{\frac{1}{2}}.
\label{Epm}
\end{align}

The dependence of the Hall conductivity on the inter-orbital pairing is now made obvious in Eq.~(\ref{eq:Sigmadp}). Apart from the energy spectra in the denominator, the inter-orbital pairing also enters through the expression $\operatorname{Re}(\Delta_{d,\vec{k}}^{\ast}\Delta_{dp,\vec{k}})$ in the numerator. The symmetry properties of the numerator fully determines whether a finite Hall conductivity may arise. If it does, it is then easy to show that for fixed $\Delta_{d}$, $\sigma_{H}^{dp} \propto \Delta_{dp}$ in the limit $|\Delta_{dp}| \ll |\Delta_{d}|$. On the side, we also note that the original $dpp$-model supports finite $\sigma_{H}$ even when only the inter-orbital pairings are present, i.e. $\Delta_{d}=0$. However, as we numerically confirmed in Fig.~\ref{fig:fig5}, $\sigma_{H} \propto |\Delta_{dp}|^2$ in that case, making this part of the contribution much smaller in comparison.

\begin{figure}
\includegraphics[width=8cm]{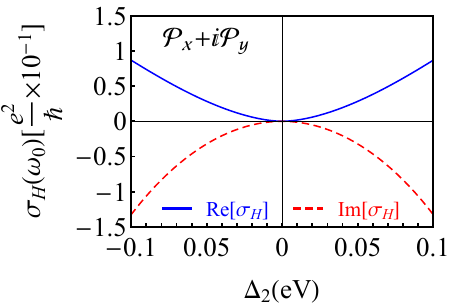}
\caption{$\Delta_2$-dependence of the Hall conductance in the $dpp$-model at $\hbar\omega_0=3.25$ eV and with $\Delta_1=0$, i.e., no intra-orbital pairing on the $d_{xy}$ orbital. For computational convenience, we have chosen $\Delta_2$ larger than the more realistic values of order $10^{-4}$ eV. The quadratic behavior is more evident for smaller $\Delta_2$.}
\label{fig:fig5}
\end{figure}

The two $dp$-submodels are related by a $C_4$ rotation, hence $\sigma_{H}^{dp_{y} }\left(  \omega\right)  =\sigma_{H}^{dp_{x}}\left(  \omega\right)$. Taken together, we have, for the original $dpp$-model in the limit $|\Delta_{dp}| \ll |\Delta_{d}|$,
\begin{equation}
\sigma_{H}\left(  \omega\right)  \approx2\sigma_{H}^{dp_{x}%
}\left(  \omega\right)  =4\sum_{\vec{k}}\frac{f_{\vec{k}}^{\left(
x\right)  }}{E_{+,\vec{k}}^{\left(  x\right)  }E_{-,\vec{k}}^{\left(
x\right)  }\left(  E_{+,\vec{k}}^{\left(  x\right)  }+E_{-,\vec{k}%
}^{\left(  x\right)  }\right)  \left[  \left(  E_{+,\vec{k}}^{\left(
x\right)  }+E_{-,\vec{k}}^{\left(  x\right)  }\right)  ^{2}-\left(
\omega+i\delta\right)  ^{2}\right]  },
\end{equation}
with $E_{\pm,\vec{k}}^{\left(  x\right)  }$ is obtained by replacing the
subscript `$p$' into `$p_{x}$' in Eq.(\ref{Epm}) and
\begin{equation}
f_{\vec{k}}^{\left(  x\right)  }=i\left[  \left(  \boldsymbol{\upsilon
}_{d,\vec{k}}-\boldsymbol{\upsilon}_{p_{x},\vec{k}}\right)
\times\boldsymbol{\upsilon}_{dp_{x},\vec{k}}\right]  _{z}\epsilon_{p_{x}%
,\vec{k}}\operatorname{Re}\left(  \Delta_{d,\vec{k}}^{\ast}%
\Delta_{dp_{x},\vec{k}}\right)  \text{.}%
\end{equation}

In the following, we provide explicit expressions of $f_{\vec{k}}^{\left(  x\right)  }$ for the three chiral superconducting states discussed in the maintext. The dependence on the inter-orbital pairing can be readily read out.

(i). $\mathcal{P}_{x}+i\mathcal{P}_{y}$: $\Delta_{d,\vec{k}}=\Delta_{1}\left(
\sin k_{x}+i\sin k_{y}\right)  \text{, }\Delta_{dp_{x},\vec{k}}%
=e^{i\phi_{1}}\Delta_{2}\cos\frac{k_y}{2}$.
\begin{equation}
f_{\vec{k}}^{\left(  x\right)  }\sim t_{dp}\Delta_{1}\Delta_{2}\left[
\left(  t_{d}-t_{p||}\right)  +2t_{d}^{\prime}\cos k_{y}\right]  \left(
2t_{p||}\cos k_{x}+2t_{p\perp}\cos k_{y}+\mu_{p}\right)  \cos\phi_{1}\sin
^{2}k_{x}\cos\frac{k_{y}}{2}.
\end{equation}
Note that $\alpha = 0$ or $\pi$ according to the GL analysis.

(ii). $\mathcal{D}_{x^2-y^2}+i\mathcal{D}_{xy}$: $\Delta_{d,\vec{k}}=\Delta
_{1}\left(  \cos k_{x}-\cos k_{y}\right)  +i\Delta_{1}^{\prime}\sin k_{x}\sin
k_{y}$, $\Delta_{dp_{x},\vec{k}}=e^{i\phi}\left(  \Delta_{2}\sin\frac
{k_{y}}{2}+i\Delta_{2}^{\prime}\sin\frac{k_{x}}{2}\right)  .$
\begin{align}
f_{\vec{k}}^{\left(  x\right)  } &  \sim t_{dp}\left[  \left(
t_{d}-t_{p||}\right)  +2t_{d}^{\prime}\cos k_{y}\right]  \left(  2t_{p||}\cos
k_{x}+2t_{p\perp}\cos k_{y}+\mu_{p}\right)  \nonumber\\
&  \text{ \ }\times\sin\phi\left[  \Delta_{1}\Delta_{2}^{\prime}\sin k_{x}%
\sin\frac{k_{x}}{2}\sin\frac{k_{y}}{2}\left(  \cos k_{x}-\cos k_{y}\right)
+\frac{1}{2}\Delta_{1}^{\prime}\Delta_{2}\sin^{2}k_{x}\sin^{2}k_{y}\right]  \,,
\end{align}
where $\phi=\pm \pi/2$.

(iii). $\mathcal{D}_{xz}+i\mathcal{D}_{yz}$: $\Delta_{d,\vec{k}}=\Delta
_{1}\left(  \sin k_{x}+i\sin k_{y}\right)  \sin k_{z}$, $\Delta_{dp_{x}%
,\vec{k}}=e^{i\phi_{1}}\Delta_{2}\cos\frac{k_{y}}{2}\sin k_{z}$.
\begin{equation}
f_{\vec{k}}^{\left(  x\right)  }\sim t_{dp}\Delta_{1}\Delta_{2}\left[
\left(  t_{d}-t_{p||}\right)  +2t_{d}^{\prime}\cos k_{y}\right]  \left(
2t_{p||}\cos k_{x}+2t_{p\perp}\cos k_{y}+\mu_{p}\right)  \cos\phi_{1}\cos
^{2}\frac{k_{y}}{2}\sin^{2}k_{z} \,,
\end{equation}
where $\alpha=0$ or $\pi$.

\subsection{D. $d^3$-model}
Here we construct the model containing three $t_{2g}$ orbitals. Since spin-orbit coupling mixes opposite spins of the $d_{xy}$ and the quasi-1D orbitals, we may take the basis $\psi_{\vec{k}\sigma}=(c_{xz\vec{k}\sigma},c_{yz\vec{k}\sigma},c_{xy\vec{k}\bar{\sigma}})^T$. Here the three components of the spinor represent the annihilation operators of the three Ru $t_{2g}$ orbitals and $\bar{\sigma}$ denotes the opposite of spin-$\sigma$. The normal state band structure is commonly described by the following Hamiltonian (for $\sigma = \ua$),
\begin{equation}
\hat{H}_{0\vec{k}}=\begin{pmatrix}
\epsilon_{1\vec{k}} & \lambda \sin k_x \sin k_y - i\eta &  i\eta \\
\lambda \sin k_x \sin k_y + i\eta & \epsilon_{2\vec{k}} & -\eta \\
-i\eta & -\eta & \epsilon_{3\vec{k}}
\end{pmatrix} \,.
\label{eq:H0_3d}
\end{equation}
Here $\epsilon_{1\vec{k}}=-2t\cos k_x-2\tilde{t}\cos k_y-\mu$, $\epsilon_{2\vec{k}}=-2\tilde{t}\cos k_x-2t\cos k_y-\mu$ and $\epsilon_{3\vec{k}}=-2t^\prime(\cos k_x+\cos k_y)-4t^{\prime\prime}\sin k_x \sin k_y-\mu_{xy}$, where $(t, \tilde{t}, t^\prime, t^{\prime\prime}, \lambda, \eta, \mu, \mu_{xy})=(0.5, 0.05, 0.4, 0.15, 0.1, 0.05, 0.45, 0.55)$ eV. Note that $\eta$ stands for the spin-orbit coupling. The other subblock of the Hamiltonian (for $\sigma=\da$) can be found in various literature, such as Ref.~\onlinecite{Scaffidi:14}.

We consider the example of chiral p-wave pairing, e.g. the pairing in the $E_u$ irrep. The primary intra-orbital chiral p-wave pairing on the $d_{xy}$ orbital is described by the same gap function as in the maintext, and the intra-orbital pairing on the $d_{xz}$ and $d_{yz}$ orbitals are considered negligible within our proposal. As for the inter-orbital pairing between $d_{xy}$ and the other two orbitals, there exists multiple $E_u$ pairings in the same symmetry channel~\cite{Huang:19b,Ramires:19,Kaba:19}. Analogous to the $dpp$-model, the inter-orbital pairing may be induced by inter-orbital mixing which couples the $d_{xy}$ orbital with the other two, such as the spin-orbit coupling. Our Ginzburg-Landau analysis reveals that not all of the inter-orbital $E_u$ pairings listed in Table II of Ref.~\onlinecite{Huang:19b} can be induced. For our purpose, we take one that can emerge due to the coupling, $(T_6\otimes k_y \vec{x}\cdot \vec{s},T_4\otimes k_x \vec{y}\cdot \vec{s})$, from the reference. This is an orbital-triplet and spin-triplet pairing. With appropriate input about the relative phases between the pairing components from the free-energy analysis, the full pairing term in the BdG Hamiltonian then reads $\sum_{\vec{k}} \psi^\dagger_{\vec{k}\sigma} \hat{\Delta}_{\vec k}\psi^\dagger_{-\vec k \bar{\sigma}} + h.c.$, where
\begin{equation}
\hat{\Delta}_{\vec{k}} =\begin{pmatrix}
0& 0 & i\Delta_2\sin k_x \\
0 & 0 & -i \Delta_2 \sin k_y \\
i \Delta_2 \sin k_x & i\Delta_2 \sin k_y & \Delta_1 (\sin k_x+i\sin k_y)
\end{pmatrix} \,,
\end{equation}
In the numerical computation presented in the maintext, we again take $\Delta_1 = 0.35$ meV and $\Delta_2=\Delta_1/10$.

Note that we have not included interlayer inter-orbital hybridization in Eq.~(\ref{eq:H0_3d}). In some pairing channels, certain inter-orbital pairings can only be induced by interlayer hybridization, instead of SOC. We will not elaborate this aspect here.

\end{widetext}


\begin{thebibliography}{99}
\bibitem{Maeno:94} Y. Maeno, H. Hashimoto, K. Yoshida, S. Nishizaki, T. Fujita, J. G. Bednorz, F. Lichtenberg, Nature (London) {\bf 372}, 532 (1994).
\bibitem{Maeno:01} Y. Maeno, T. M. Rice, and M. Sigrist, Phys. Today {\bf 54}(1), 42 (2001).
\bibitem{Mackenzie:03} A. P. Mackenzie and Y. Maeno, Rev. Mod. Phys. {\bf 75}, 657 (2003).
\bibitem{Kallin:09} C. Kallin and A. J. Berlinsky, J. Phys. Condens. Matter {\bf 21}, 164210 (2009).
\bibitem{Kallin:12} C. Kallin, Rep. Prog. Phys. {\bf 75}, 042501 (2012).
\bibitem{Maeno:12} Y. Maeno, S. Kittaka, T. Nomura, S. Yonezawa, and K. Ishida, J. Phys. Soc. Jpn. {\bf 81}, 011009 (2012).
\bibitem{Liu:15} Y. Liu and Z. Q. Mao, Physica C: Superconductivity and its Application, {\bf 514}, 339 (2015).
\bibitem{Kallin:16} C. Kallin and A. J. Berlinsky, Rep. Prog. Phys. {\bf 79}, 054502 (2016).
\bibitem{Mackenzie:17} A. P. Mackenzie, T. Scaffidi, C. W. Hicks and Y. Maeno, NPJ Quantum Materials {\bf 2}, 40 (2017).
\bibitem{Luke:98} G. M. Luke, Y. Fudamoto, K. M. Kojima, M. I. Larkin, J. Merrin, B. Nachumi, Y. J. Uemura, Y. Maeno, Z. Q. Mao, Y. Mori, H. Nakamura, M. Sigrist, Nature {\bf 394}, 558 (1998).
\bibitem{Xia:06} J. Xia, Y. Maeno, P. T. Beyersdorf, M. M. Fejer, and A. Kapitulnik, Phys. Rev. Lett. {\bf 97}, 167002 (2006).
\bibitem{Grinenko:20} V. Grinenko, S. Ghosh, R. Sarkar, J. Orain, A. Nikitin, M. Elender, D. Das, Z. Guguchia, F. Br\"uckner, M. E. Barber, J. Park, N. Kikugawa, D. A. Sokolov, J. S. Bobowski, T. Miyoshi, Y. Maeno, A. P. Mackenzie, H. Luetkens, C. W. Hicks, H. Klauss, arXiv:2001.08152.
\bibitem{Kitaev:03} A. Yu. Kitaev, Annals. Phys. {\bf 303}, 2 (2003).
\bibitem{Nayak:08} C. Nayak, S. H. Simon, A. Stern, M. Freedman, S. Das Sarma, Rev. Mod. Phys. {\bf 80}, 1083 (2008).
\bibitem{Read:00} N. Read and D. Green, Phys. Rev. {\bf B 61}, 10267 (2000).
\bibitem{Yip:92} S. K. Yip and J. A. Sauls, J. Low Temp. Phys. {\bf 86}, 257 (1992).
\bibitem{Roy:08} R. Roy and C. Kallin, Phys. Rev. B {\bf 77}, 174513 (2008).
\bibitem{Lutchyn:08} R. M. Lutchyn, P. Nagornykh, and V. M. Yakovenko, Phys. Rev. B 77, 144516 (2008).
\bibitem{Goryo:08} J. Goryo, Phys. Rev. B {\bf 78}, 060501(R) (2008).
\bibitem{Lutchyn:09} R. M. Lutchyn, P. Nagornykh, and V. M. Yakovenko, Phys. Rev. B {\bf 80}, 104508 (2009).
\bibitem{Konig:17} E. J. Konig, A. Levchenko, Phys. Rev. Lett. {\bf 118}, 027001 (2017).
\bibitem{LiYu:19} Y. Li, Z. Wang, and W. Huang, arXiv:1909.08012.
\bibitem{Taylor:12} E. Taylor and C. Kallin, Phys. Rev. Lett. {\bf 108}, 157001 (2012).
\bibitem{Wysokinski:12} K.I. Wysoki\'nski, J. F. Annett, B. L. Gy\"orffy, Phys. Rev. Lett. {\bf 108}, 077004 (2012).
\bibitem{Komendova:17} L. Komendova and A. M. Black-Schaffer, Phys. Rev. Lett. {\bf 119}, 087001 (2017).
\bibitem{Vaugier:12}  L. Vaugier, H. Jiang, and S. Biermann. Phys. Rev. B {\bf86}, 165105 (2012).
\bibitem{Oguchi:95} T. Oguchi. Phys. Rev. B {\bf 51}, 1385 (1995).
\bibitem{Supplementary} See Supplemental Materials.
\bibitem{footnote1} Infinitesmall intra-orbital pairing on orbitals far from Fermi level can in principle be induced by the dominant superconducting orbitals through inter-orbital mixing. But we neglect it in the present study. Same with the $d^3$-model where we ignore intra-orbital pairing on the quasi-1D $d$ orbitals.
\bibitem{Huang:18} W. Huang and H. Yao, Phys. Rev. Lett. {\bf 121}, 157002 (2018).
\bibitem{Huang:19b} W. Huang, Y. Zhou, and H. Yao, Phys. Rev. B {\bf 100}, 134506 (2019).
\bibitem{Ramires:19} A. Ramires and M. Sigrist, Phys. Rev. B {\bf 100}, 104501 (2019).
\bibitem{Kaba:19} S.O. Kaba, D. S\'en\'echal, Phys. Rev. B {\bf 100}, 214507 (2019).
\bibitem{Gradhand:13} M. Gradhand, K. Wysoki\'nski, J. Annett, B Gy\"orffy. Phys. Rev. B {\bf88}, 094504 (2013).
\bibitem{Sharma:20} R. Sharma, S. Edkins, Z. Wang, A. Kostin, {\it et al.}, PNAS {\bf 117}, 5222 (2020).
\bibitem{Yoshioka:18} N. Yoshioka, Y. Imai and M. Sigrist, J. Phys. Soc. Jpn. {87}, 124602 (2018).
\bibitem{Nie:19} W. Nie, W. Huang, and H. Yao, Phys. Rev. B {\bf 102}, 054502 (2020).
\bibitem{Wang:18} Z. Q. Wang, J. Berlinsky, G. Zwicknagl, and C. Kallin, Phys. Rev. B {\bf 96}, 174511 (2017).
\bibitem{Brydon:19} P. M. R. Brydon, D. S. L. Abergel, D. F. Agterberg, and V. M. Yakovenko, Phys. Rev. X {\bf 9}, 031025 (2019).
\bibitem{Bergemann:03} C. Bergemann, A. P. Mackenzie, S.R. Julian, D. Forsythe, and E. Ohmichi, Adv. Phys. {\bf 52}, 639 (2003).
\bibitem{Haverkort:08} M. W. Haverkort, I. S. Elfimov, L. H. Tjeng, G. A. Sawatzky, and A. Damascelli, Phys. Rev. Lett. {\bf 101}, 026406 (2008).
\bibitem{Veenstra:14} C. N. Veenstra, Z. -H. Zhu, M. Raichle, B. M. Ludbrook, A. Nicolaou, B. Slomski, G. Landolt, S. Kittaka, Y. Maeno, J. H. Dil, I. S. Elfimov, M. W. Haverkort, and A. Damascelli, Phys. Rev. Lett. {\bf 112}, 127002 (2014).
\bibitem{Gingras:19} O. Gingras, R. Nourafkan, A. -M. S. Tremblay, and M. C\`ot\'e, Phys. Rev. Lett. {\bf 123}, 217005 (2019).
\bibitem{Pustogow:19} A. Pustogow, Y. Luo, A. Chronister, Y. Su, D. A. Sokolov, F. Jerzembeck, A. P. Mackenzie, C. W. Hicks, N. Kikugawa, S. Raghu, E. D. Bauer, S. E. Brown, Nature (2019).
\bibitem{Roising:19} H. R\o ising, T. Scaffidi, F. Flicker, G. Lange, and S. Simon. Phys. Rev. Research {\bf 1}, 033108 (2019).
\bibitem{Romer:19} A. T. R\o mer, D. D. Scherer, I. M. Eremin, P. J. Hirschfeld, and B. M. Andersen, Phys. Rev. Lett. {\bf 123}, 247001 (2019).
\bibitem{Romer:20} A. T. R\o mer, A. Kreisel, M. M\"uller, P. J. Hirschfeld, I. M. Eremin, B. M. Andersen, arXiv:2003.13340.
\bibitem{Kivelson:20} S. A. Kivelson, A. C. Yuan, B. J. Ramshaw, R. Thomale, arXiv:2002.00016.
\bibitem{Chronister:20} A. Chronister, A. Pustogow, N. Kikugawa, D. A. Sokolov, F. Jerzembeck, C.W. Hicks, A. P. Mackenzie, E. D. Bauer, S.E. Brown, arXiv:2007.13730.
\bibitem{Scaffidi:20} T. Scaffidi, arXiv:2007.13769.
\bibitem{Huang:15} W. Huang, S. Lederer, E. Taylor, C. Kallin, Phys. Rev. B {\bf 91}, 094507 (2015).
\bibitem{Kirtley:07} J. R. Kirtley, C. Kallin, C. W. Hicks, E. -A. Kim, Y. Liu, K. A. Moler, Y. Maeno, K. D. Nelson, Phys. Rev. B {\bf 76}, 014526 (2007).
\bibitem{Hicks:10} C. W. Hicks, J. R. Kirtley, T. M. Lippman, N. C. Koshnick, M. E. Huber, Y. Maeno, W. M. Yuhasz, M. B. Maple, K. A. Moler, Phys. Rev. B {\bf 81}, 214501 (2010).
\bibitem{Curran:14} P. J. Curran, S. J. Bending, W. M. Desoky, A. S. Gibbs, S. L. Lee,and  A. P. Mackenzie, Phys. Rev. B {\bf 89}, 144504 (2014).
\bibitem{Huang:14} W. Huang, E. Taylor, and C. Kallin, Phys. Rev. B {\bf 90}, 224519 (2014).
\bibitem{Tada:15} Y. Tada, W. Nie, and M. Oshikawa, Phys. Rev. Lett. {\bf 114}, 195301 (2015).
\bibitem{Yonezawa:14} S. Yonezawa, T. Kajikawa, Y. Maeno, J. Phys. Soc. Jpn. {\bf 83}, 083706 (2014).
\bibitem{Li:19} Y. -S. Li, N. Kikugawa, D. A. Sokolov, F. Jerzembeck, A. S. Gibbs, Y. Maeno, C. W. Hicks, M. Nicklas, A.P. Mackenzie, arXiv:1906.07597.
\bibitem{Ghosh:20} S. Ghosh, A. Shekhter, F. Jerzembeck, N. Kikugawa, D. A. Sokolov, M. Brando, A. P. Mackenzie, C. W. Hicks, B. J. Ramshaw, arXiv:2002.06130.
\bibitem{Ishida:19} K. Ishida, M. Manago, and Y. Maeno, arXiv:1907.12236 (2019).
\bibitem{Petsch:20} A. N. Petsch, M. Zhu, Mechthild Enderle, Z. Q. Mao, Y. Maeno, S. M. Hayden, arXiv:2002.02856.
\bibitem{Benhabib:20} S. Benhabib, C. Lupien, I. Paul, L. Berges, M. Dion, M. Nardone, A. Zitouni, Z.Q. Mao, Y. Maeno, A. Georges, L. Taillefer, C. Proust, arXiv:2002.05916.
\bibitem{Agterberg:97} D. F. Agterberg, T. M. Rice, and M. Sigrist, Phys. Rev. Lett. {\bf 78}, 3374 (1997).
\bibitem{Mazin:99} I.I. Mazin and D.J. Singh, Phys. Rev. Lett. {\bf 82}, 4324 (1999).
\bibitem{Nomura:00} T. Nomura and K. Yamada, J. Phys. Soc. Jpn. {\bf 69}, 3678 (2000).
\bibitem{Nomura:02} T. Nomura and K. Yamada, J. Phys. Soc. Jpn. {\bf 71}, 1993 (2002).
\bibitem{Zhitomirsky:01} M. E. Zhitomirsky, and T. M. Rice, Phys. Rev. Lett. {\bf 87}, 057001 (2001).
\bibitem{Ng:00} K. K. Ng and M. Sigrist, EPL {\bf 49}, 473 (2000).
\bibitem{Eremin:02} I. Eremin, D. Manske and K. H. Bennemann, Phys. Rev. B {\bf 65}, 220502(R) (2002).
\bibitem{RPA1} T. Takimoto, T. Hotta, and K. Ueda, Phys. Rev. B \textbf{69}, 104504 (2004).
\bibitem{RPA2} K. Yada and H. Kontani, J. Phys. Soc. Jpn. \textbf{74}, 2161 (2005).
\bibitem{Annett:06} J. F. Annett, G. Litak, B. L. Gy\"orffy, and K. I. Wysoki\'nski, Phys. Rev. B 73, 134501 (2006).
\bibitem{RPA3} K. Kubo, Phys. Rev. B \textbf{75}, 224509 (2007).
\bibitem{Raghu:10} S. Raghu, A. Kapitulnik, S. A. Kivelson, Phys. Rev. Lett. {\bf 105}, 136401 (2010).
\bibitem{Puetter:12} C. M. Puetter and H. -Y. Kee, Europhys. Lett. {\bf 98}, 27010 (2012).
\bibitem{Huo:13} J. W. Huo, T. M. Rice and F. C. Zhang, Phys. Rev. Lett. {\bf 110}, 167003 (2013).
\bibitem{Wang:13} Q. H. Wang, C. Platt, Y. Yang, C. Honerkamp, F. C. Zhang, W. Hanke, T. M. Rice and R. Thomale, Europhys. Lett. {\bf 104}, 17013 (2013).
\bibitem{Kivelson:13} I. A. Firmo, S. Lederer, C. Lupien, A. P. Mackenzie, J. C. Davis, and S. A. Kivelson, Phys. Rev. B {\bf 88}, 134521 (2013).
\bibitem{Yanase:14} Y. Yanase, S. Takamatsu, and M. Udagawa, J. Phys. Soc. Jpn. {\bf 83}, 061019 (2014).
\bibitem{Scaffidi:14} T. Scaffidi, J. C. Romers, and S. H. Simon, Phys. Rev. B {\bf 89}, 220510(R) (2014).
\bibitem{Tsuchiizu:15}  M. Tsuchiizu, Y. Yamakawa, S. Onari, Y. Ohno, and H. Kontani, Phys. Rev. B {\bf 91}, 155103 (2015).
\bibitem{Huang:16} W. Huang, T. Scaffidi, M. Sigrist, and C. Kallin, Phys. Rev. B {\bf 94}, 064508 (2016).
\bibitem{Ramires:16} A. Ramires and M. Sigrist, Phys. Rev. B {\bf 94}, 104501 (2016).
\bibitem{ZhangLD:18} L.-D. Zhang, W. Huang, F. Yang, and H. Yao, Phys. Rev. B {\bf 97}, 060510 (2018).
\bibitem{WangWS:19} W. -S. Wang, C. -C. Zhang, F. -C. Zhang, and Q. -H. Wang, Phys. Rev. Lett. {\bf 122}, 027002 (2019).
\bibitem{WangZQ:20} Z. Wang, X. Wang, and C. Kallin, Phys. Rev. B {\bf 101}, 064507 (2020).
\bibitem{Suh:19} H. G. Suh, H. Menke, P. M. R. Brydon, C. Timm, A. Ramires, D. F. Agterberg, arXiv:1912.09525.
\bibitem{Lindquist:19} A. W. Lindquist, H-Y. Kee, arXiv:1912.02215.
\bibitem{ChenW:20} W. Chen and J. An, arXiv:2004.04941.
\bibitem{Schemm:14} E. R. Schemm, W. J. Gannon, C. M. Wishne, W. P. Halperin, A. Kapitulnik, Science {\bf 345}, 190 (2014).
\bibitem{Schemm:15} E. R. Schemm, R. E. Baumbach, P. H. Tobash, F. Ronning, E. D. Bauer, A. Kapitulnik, Phys. Rev. B {\bf 91}, 140506(R) (2015).
\bibitem{Levenson:18} E. M. Levenson-Falk, E.R. Schemm, Y. Aoki, M. B. Maple, A. Kapitulnik, Phys. Rev. Lett. {\bf 120}, 187004. (2018).
\bibitem{Hayes:20}  I.M. Hayes, D.S. Wei, T. Metz, J. Zhang, Y.S. Eo, S. Ran, S.R. Saha, J. Collini, N.P. Butch, D.F. Agterberg, A. Kapitulnik, and J. Paglione, arXiv:2002.02539.

\end{thebibliography}
\end{document}